\documentclass[pra,twocolumn,psfig]{revtex4}
\def\bp{\beta^\prime}
\def\bdp{\beta^{\prime\prime}}
\usepackage{psfig}
\usepackage{bm}

\newcommand{\uvec}{\bm u}
\newcommand{\nuvec}{\bm\nu}

\newcommand{\xvec}{\bm x}
\newcommand{\omvec}{\bm\omega}
\newcommand{\Omvec}{\bm\Omega}
\newcommand{\xivec}{\bm\xi}
\newcommand{\grad}{\bm\nabla}
\newcommand{\bnabla}{\bm\nabla}
\newcommand{\curl}{\grad\crossprod}
\def\be{\begin{equation}}
\def\ee{\end{equation}}
\def\baray{\begin{eqnarray}}
\def\earay{\end{eqnarray}}

\newcommand{\crossprod}{\bm \times}

\begin{document}
\title{Oscillations of Bose-Einstein condensates with vortex
lattices:\\
  Finite temperatures}
\author{Armen Sedrakian$^1$ and Ira Wasserman$^2$}
\affiliation{$^1$ Institute for Theoretical Physics, 
T\"ubingen University, D-72076 T\"ubingen, Germany\\
$^2$ Center for Radiophysics and Space Research and Newman Laboratory, \\
Cornell University, 
Ithaca, New York 14853, USA 
}

\begin{abstract}
We derive the finite temperature oscillation modes of a harmonically 
confined Bose-Einstein condensed gas undergoing rigid body rotation 
supported by a vortex lattice in the condensate. 
The hydrodynamic modes separate into two classes
corresponding to center-of-mass and
relative oscillations of the thermal cloud and the condensate. 
These classes are independent of each other 
in the case where the thermal cloud is inviscid for all modes studied, 
except the radial pulsations which couple them because the pressure
perturbations of the condensate and the thermal cloud are governed by 
different adiabatic indices. If the thermal cloud is viscous, the 
two classes of oscillations are coupled, i.e., each type of motion 
involves simultaneously mass and entropy currents. 
The relative oscillations are damped by the 
mutual friction between the condensate and the thermal cloud 
mediated by the vortex lattice. The damping is large for the
values of the drag-to-lift ratio of the order of unity and becomes 
increasingly ineffective in either limit of small or large friction.
An experimental measurement of a subset of these  oscillation modes  
and their damping  can provide information on the values of 
the phenomenological mutual friction coefficients and the 
quasiparticle-vortex scattering processes in dilute atomic Bose gases.
\end{abstract}

%\maketitle

\maketitle

\section{Introduction}
Recent experiments on rotating trapped Bose-Einstein condensates
have created vortex lattices with large number of 
vortices either by stirring a condensate with a rotating anisotropic 
perturbation \cite{PARIS1,MIT1,MIT2,OXFORD} or by 
evaporatively cooling a spinning normal gas
\cite{JILA1}.  Being reminiscent of
the classical rotating bucket experiments,  carried out on
superfluid phases of helium, the experiments on rotating 
Bose-Einstein condensates open a variety of new ways of 
manipulating a rotating condensate and deducing information on 
its physical properties \cite{PARIS1,MIT1,MIT2,OXFORD,JILA1}.
The aspect ratios of the condensates measured by a non destructive imaging
and the deduced number of vortices are consistent
with a rigid body rotation of the condensates.
The experiments carried out at finite temperatures \cite{MIT1,JILA1}
achieve a rigid body rotation of the thermal cloud
at a   frequency close to that of the condensate (for example, 
in the MIT experiment \cite{MIT1} the ratio of spin frequencies 
of the thermal cloud to that of the condensate is close to  2/3). 
These systems contain typically $10^5$-$10^6$ particles and 
the fluid hydrodynamics is well suited for the studies of their confined 
state. The range of rotation rates covered by the experiments extends up to 
the centrifugal limit: by combining the evaporative cooling and
optical spin-up techniques rotation rates larger than 99$\%$ of the
centrifugal limit  were achieved for a harmonic confinement \cite{JILA2}; 
rotation rates beyond the centrifugal limit were reached by
a sequential laser stirring of a condensate confined in a 
combination of a quadratic and quartic potentials \cite{PARIS2}.

In this paper we address the  oscillations and stability of a
harmonically confined, 
rotating Bose-Einstein condesate at finite temperatures in the hydrodynamic
regime. The angular momentum of the superfluid is carried by singly 
quantized vortices which form a triangular Abrikosov lattice. We work 
in the limit of coarse-grained hydrodynamics, where the physical quantities
are averages over large number of vortices. In this limit the
structure of individual vortices is unresolved and the superfluid
simply mimics a rigid body rotation at the frequency $\Omega = \kappa
n_V/2$, where $\kappa = 2\pi\hbar/m$ is the quantum of circulation, 
$n_V$ is the vortex density in the plane orthogonal to the spin
vector, $m$ is the boson mass.
The hydrodynamic approximation to the microscopic dynamics of
the Bose-Einstein condesate is justified in the regime of strong
interparticle interactions and large number of particle in the system
(the strong-coupling regime requires $Na/d\gg 1$
where $N$ is the net number of particles in the condensate, $a$ is the
scattering length, and $d=\sqrt{\hbar/m\omega_0}$ is the oscillator
length defined in terms of oscillator frequency $\omega_0$). 
The Thomas-Fermi approximation is valid 
for systems with large number of particles and insures that quantum
corrections (such as the `quantum pressure' term) to the hydrodynamic
equations can be neglected. In addition, for large enough systems 
the space-local values of the thermodynamic quantities such as the 
pressure and chemical potential are well defined. For such systems the
coherence length of the condensate $\xi = 1/\sqrt{8\pi na}$, 
where $n$ is the number density of the condensate particles, 
becomes much smaller than the intervortex distance and the 
vortex cores can be treated as singularities in the hydrodynamic equations.

Compared to the zero-temperature case the number of degrees of freedom 
of a Bose-Einstein condesate is doubled at finite temperatures and so is the 
number of oscillations modes. The modes can be classified into two
subsets, which correspond to the center of mass and relative
oscillations of the condensate and the thermal cloud. 
The first set, which corresponds to density oscillations of the 
combined fluid (first sound), is identical to the modes of a rotating,
zero-temperature condensate in the limit of inviscid normal-fluid
and for those modes which do not involve pressure perturbations.
These modes were derived within the tensor virial 
method \cite{CHANDRA} in a preceding paper
~\cite{Paper1} (hereafter Paper I). In the special case of  
an axial-symmetric trap, the lowest-order non-trivial modes
(classified by corresponding terms of the expansion of perturbations
in spherical harmonics labeled by indices $l,m$) are those with $l=2$ and 
$-2\le m\le 2$. Simple analytical results are available in the case
of an axisymmetric trap and we list them below for later
references. An analysis of the hydrodynamic modes 
in Ref. \cite{STRINGARI} of the even $m$ modes belonging to $l=2$ harmonics 
is in agreement with the results quoted below.  A  generalization to 
odd $m$ and arbitrary $l\ge 2$ harmonics of the surface modes of a rotating
condensate is given in Ref. \cite{CHEVY}. 
The frequencies of the {\it toroidal modes} are given by 
[Paper I, Eq. (34); our nomenclature follows Ref. \cite{CHANDRA}]
\baray\label{int1}
\sigma_{1,2}(l=2,\vert m\vert =2) = \Omega\pm\sqrt{2\omega_{\perp}^2-\Omega^2}, 
\earay
where $\Omega$ is spin frequency of the condensate, $\omega_{\perp}$ is the
component of the trapping frequency orthogonal to the spin vector;
two complementary modes follow from the substitution $\Omega \to
-\Omega$. The  frequencies of the {\it pulsation, or breathing, modes} 
are given by [Paper I, Eq. (45)]
\baray\label{int2}
\sigma_{1,2}^2(l=2,\vert m\vert =0) &=&\frac{3}{2}\omega_z^2+2\omega_{\perp}^2
\nonumber\\
&&\hspace{-2.cm}\pm\sqrt{9\omega_z^4-16(\omega_z^2\omega_{\perp}^2-\omega_{\perp}^4) -8\omega_z^2\Omega^2}, 
\earay
where $\omega_z$ is the component of the trapping frequency 
parallel to the spin vector. Finally, the frequencies of the
{\it transverse-shear modes} $(l=2,\vert m\vert =1)$  are determined 
by the characteristic equation [Paper I, Eq. (22)]
\be\label{int3}
\sigma^3-2\Omega\sigma^2 - (\omega_{\perp}^2+\omega_z^2-\Omega^2)\sigma +2\Omega\omega_z^2
= 0.
\ee
There are three distinct modes that solve  Eq. (\ref{int3}) 
\baray\label{int4}
\sigma_{1} &=& \frac{2\Omega}{3} +(s_++s_-), \nonumber\\
\sigma_{2,3} &=& \frac{2\Omega}{3} -\frac{1}{2}(s_++s_-)
\pm \frac{i\sqrt{3}}{2}(s_+-s_-),
\earay
and three complementary modes follow from the substitution $\Omega \to
-\Omega$ [see Paper I, Eq. (23) and (24)].
The coefficients in Eqs. (\ref{int4}) are defined as \cite{FOOTNOTE}
\baray\label{int4bis}
s_{\pm}^3 &=& \frac{\Omega}{3}\left(
\omega_{\perp} ^2 -2 \omega_z^2 -\frac{\Omega^2}{9}
\right)\mp \frac{1}{9}\Bigl[
\left(\omega_{\perp}^2+\omega_z^2+\frac{\Omega^2}{3}\right)^3 \nonumber\\
&&-\Omega^2
\left(\omega_{\perp}^2-2\omega_z^2-\frac{\Omega^2}{9}\right)^2
\Bigr]^{1/2}.
\earay
As we shall see below,  Eqs. (\ref{int1}) and (\ref{int3})-(\ref{int4bis})
remain valid at finite temperatures when the viscosity of the thermal cloud is
negligible;  Eq. (\ref{int2}) will be modified since the pulsations of
the condensate and the thermal cloud couple due to the difference in
the underlying equations of state of these components. In the
two fluid setting these equations correspond to the first class of the
center of mass oscillation modes. 

Note that the oscillation modes  
quoted above remain valid for nonsuperfluid Bose or Fermi systems 
and superfluid Fermi systems at zero-temperature. The reason is that 
the modes are uniquely determined by the assumptions of uniform
rotation and harmonic trapping, and by the  hydrodynamic equations
of motion which are identical for nonsuperfluid Bose and Fermi fluids 
and a superfluid Fermi liquid at zero-temperature. [The  pulsations of
these systems will differ because their equations of 
states differ, but the corresponding modes can be derived
without specifying the value of the adiabatic index, see Eq. (45) in
Paper I which applies for an arbitrary adiabatic index $\gamma \neq 1$;
the case $\gamma =1$  is treated below.]

The purpose of this work is to derive the second class 
of modes which  correspond to the relative oscillations of the 
condensate and thermal  cloud under uniform rotation. We also study in
some detail the center of mass modes when these are coupled to the
relative modes of oscillations. As in the preceding paper we shall 
use the tensor virial method (Ref. \cite{CHANDRA}, and references
therein), however the underlying hydrodynamic equations are now those of 
the two fluid superfluid hydrodynamics \cite{KHALATNIKOV,PETHICK}.
Although the tensor virial method was originally
developed for the study of equilibrium and stability of uniform, incompressible, 
rotating liquid masses bound by self-gravitation \cite{CHANDRA} it
proved useful for studies of rotating Bose gases confined by harmonic
magnetic traps. The method was extended to non-uniform
compressible flows for gases with polytropic equations of state
$p\propto \rho^{\gamma}$, where $p$ is the pressure, $\rho$ is the
density, and $\gamma$ is the adiabatic index. In this (non-uniform) case 
the equilibrium figures are {\it  heterogeneous ellipsoids} of
revolution, i.e., ellipsoids with constant density surfaces being similar and
concentric to the bounding surface.

This paper is organized as follows. Section II contains the fluid
perturbation which are derived by taking the Eulerian variations of 
various moments of the hydrodynamic equations of rotating superfluids.
(Occasionally, we refer to the condensate and the thermal cloud as 
the superfluid and normal-fluid, respectively.) In Sec. III we derive 
the small-amplitude first and second-order harmonic oscillation modes 
when the normal-fluid is inviscid.  
The kinematic viscosity of the normal could is included in the virial 
equations and the numerical solutions of the corresponding
characteristic equations are presented. Our conclusions are summarized  
in Sec. IV. Appendix A contains the virial equations for uniform
rotations; Appendix B presents the Eulerian variations of the
stress energy and pressure tensors.

\section{Virial equations and their perturbations}
\subsection{Hydrodynamics in a trap}

Consider a rotating cloud of Bose condensed gas confined in a harmonic
trap. The trapping potential is characterized in terms 
of Cartesian frequency components $\omega_i$ as
\be
\phi_{\rm tr} (\xvec) = \omega_i^2x_i^2,
\ee
where we assume that
the rotation axis is along the positive $z$ direction of the Cartesian
system of coordinates, $\omega_1 = \omega_2 =\omega_{\perp}$
for axisymmetric traps, and 
we assume implicit summation over the repeated Latin 
indices from 1 to 3, unless stated otherwise.
The Euler and Navier-Stokes equations for the condensate and the
thermal cloud can be combined in a single equation 
(which is written below in a frame rotating with angular velocity 
$\Omvec$ relative to some inertial coordinate reference system)
\baray
\rho_\alpha\left({\partial\over\partial t}+u_{\alpha,j}
{\partial\over\partial x_j}\right)  u_{\alpha,i}
&=&-{\partial p_\alpha\over
\partial x_i}-\frac{\rho_{\alpha}}{2}{\partial\phi_{\rm tr}\over\partial x_i}
\nonumber\\
&&\hspace{-4.cm}
+\delta_{\alpha N}\frac{\partial P_{ik}}{\partial x_k}
+\frac{\rho_\alpha}{2}{\partial\vert\Omvec\crossprod\xvec\vert^2
\over\partial x_i}
+2\rho_\alpha\epsilon_{ilm}u_{\alpha,l}\Omega_m
+F_{\alpha\beta,i},\nonumber\\
\label{eq:euler}
\earay
where the Greek subscripts $\alpha,\beta, \dots  \in\{S,N\}$ identify
the fluid component ($S$ refers to superfluid, $N$ - to the normal-fluid);
the Latin subscripts denote the coordinate directions; $\rho$,
$p$, and $u_i$ are the density, pressure, and velocity of 
the condensate,  $P_{ik}$ is the stress tensor, and
$F_{\alpha\beta,i}$ is the mutual friction force on fluid 
$\alpha$ due to fluid $\beta$. Below, we shall assume that 
the condensate and the thermal cloud are isothermal in the 
background equilibrium, while the perturbations from the 
equilibrium state are adiabatic. 

The equation of state of the condensate, to the leading order
in diluteness parameter $n_Sa^3$, where $n_S$ is the number density
of condensate atoms, $a$ is the scattering length, can be written in a
polytropic form $p_S  = K_S\rho_S^{\gamma}$, 
where $K_S$ is a constant and $\gamma = 2$
is the adiabatic index \cite{Paper1}. The density profile of the condensate in 
a rotating trap is obtained by integrating Eq. (\ref{eq:euler}) 
in the stationary (time independent) limit with respect to the spatial
coordinates [Paper I, Eq.~(14)]
\baray\label{profile_S} 
\rho_S(\xvec) &=& \rho_S(0)
\Bigl[1- \frac{\gamma-1}{2K_S\gamma\rho_{S}(0)^{\gamma-1}}\nonumber\\
&\times&\left(\phi_{\rm tr}(\xvec)-\vert
\Omvec\crossprod\xvec\vert^2\right)\Bigr]^{1/(\gamma -1)}\theta ,
\earay
where $\theta =1 $ when the expression in the brackets is positive
and zero otherwise. While  Eq. (\ref{profile_S}) applies for 
arbitrary values of the adiabatic index, it is not valid 
in the special case $\gamma = 1$ occurring for the normal gas
in the classical limit.

The equation of state of the thermal cloud  is $p_N (\xvec) 
=\left[g_{5/2}(z))/g_{3/2}(z)\right] k_B T \rho_N(\xvec)$,
where $z = {\rm exp}(\mu/k_BT)$ is the fugacity, $T$ is the 
temperature, $k_B$ is the Boltzmann constant, $\mu$ is the chemical potential, 
and $g_{n}(z) = \sum_{l = 1}^{\infty}z^l/l^n$;
the thermodynamic quantities above refer to local equilibrium 
values. The density profile of the thermal cloud can be obtained
analytically in the temperature and density regime where the 
quantum degeneracy of the thermal cloud can be neglected. 
Upon taking the classical limit $z \to 1$ in 
the expression for $p_N(\xvec)$,  
integrating over spatial coordinates, we find the density profile of
the thermal cloud, which is Gaussian
\be\label{profile_N} 
\rho_N (\xvec) = \rho_N(0){\rm exp}
~\left[-\frac{1}{2K_N}
\left(\phi_{\rm tr}(\xvec)-\vert
\Omvec\crossprod\xvec\vert^2 \right)\right];
\ee
here $K_N = m/k_BT$. A common feature of the 
density profiles (\ref{profile_S}) and (\ref{profile_N})
is that the centrifugal and trapping potentials are 
quadratic forms of the coordinates,  $\rho_{S,N} 
= \rho_{S,N}(m_{S,N}^2)$, where
\be
m_{S,N}^2 = \frac{x_1^2}{a_{1,N,S}^2}+\frac{x_2^2}{a_{2,N,S}^2}
+\frac{x_3^2}{a_{3,N,S}^2},\quad m_{S,N}\le  1.
\ee
As demonstrated in Appendix A,  associated with these density 
distributions are heterogeneous ellipsoids of the condensate and the thermal cloud 
with semi-major axis $a_{i,N,S}$ ($i = 1,2,3$). This observation
serves as a starting point for the application of the tensor virial
method to describe the equilibrium and stability of ellipsoidal
figures of Bose condensed gases (see also Paper I).

Because of the difference in the equations of state of the 
condensate and the thermal cloud the underlying hydrodynamical
equations [i.e., the 
components of Eq. (\ref{eq:euler})] are not invariant with 
respect to an interchange $\alpha\leftrightarrow\beta$ 
of the indices labeling the dynamical components. As a result, 
the perturbed motions of the components cannot be separated into purely 
center of mass and relative oscillations for the modes that involve
pressure perturbations. This symmetry is 
also broken because of the viscosity of the thermal cloud. 
The Navier-Stokes equation for the normal-fluid 
contains the stress tensor in its common form
\be\label{eq:stress_tensor} 
P_{ik}=\rho_N\nu\left(\frac{\partial u_{N,i}}{\partial x_k}
+\frac{\partial u_{N,k}}{\partial x_i}
-\frac{2}{3}\frac{\partial  u_{N,l}}{\partial x_l}\delta_{ik}\right),
\ee
where $\nu$ is the kinematic viscosity. 
In the inviscid limit and for the modes that do not involve pressure
perturbations the center of mass and relative oscillations are
decoupled due to the symmetry above. And these oscillations are coupled 
whenever the modes require pressure perturbations (e.g., the pulsation modes)
or the viscosity of the thermal cloud is operative.

Consider the condensate and the thermal cloud  rotating uniformly 
at the same spin frequency; (i.e., we assume the 
external torque is time independent and/or the cloud
and condensate had sufficient 
time to relax to a rotational equilibrium).
When perturbed from equilibrium the fluids interact via 
the mutual friction force:
\be\label{eq:MF}
F_{\alpha\beta,i}=-{\cal S}_{\alpha\beta}
\rho_S\omega_S\beta_{ij}(u_{S,j}-u_{N,j}),
\ee
where the mutual friction tensor is
$
\beta_{ij}=\beta\delta_{ij}+\beta^\prime\epsilon_{ijm}\nu_m
+(\bdp-\beta)\nu_i\nu_j,
$
with $\beta$, $\bp$ and $\bdp$ being the mutual friction
coefficients, $\omvec_{S}=\nuvec\omega_S\equiv\curl\uvec_S$
and ${S}_{\alpha\beta}$ is a fully antisymmetric (second rank)
unit tensor  with the sign convention ${\cal S}_{SN} =1$. 
Note that Eq.  (\ref{eq:MF}) is the local form of the mutual friction;
when constructing the global (integrated over volume) virial equation 
we need to take into account the fact that the force acts only 
within the combined volume of the two fluids.
Equation (\ref{eq:MF}) can be put in a form reflecting the
balance of forces acting on a vortex
\begin{eqnarray}\label{eq:AS2.4.2}
\rho_S\omega_S\epsilon_{ijm}(u_{Sj}-u_{Lj}) \nu_m
-\eta_{ij}(u_{Lj}-u_{nj}) =0,
\end{eqnarray}
where $\eta_{ij} = \eta\delta_{ij}+\eta'\epsilon_{ijm}\nu_m$,
and the components of the friction tensors are related by
\begin{eqnarray}\label{eq:AS:etas}
\beta = \frac{\zeta(1+\zeta')}{\zeta^2+1},\quad
\beta' = \frac{\zeta^2-\zeta'}
{\zeta^2+1},
\end{eqnarray}
where $\zeta = \eta/(\rho_S\kappa-\eta')$
$\zeta' = \eta'/(\rho_S\kappa-\eta')$ are the (dimensionless) 
drag-to-lift ratios.  
The first term in Eq. (\ref{eq:AS2.4.2}) is a
nondissipative lifting force due to a superflow past the vortex
(the Magnus force).
The remaining terms reflect the friction between the vortex
and the normal-fluid. 
Note that, by definition, work is done only by the component of 
the friction force $\propto \eta$, which is  parallel to 
the vortex motion; the orthogonal component $\propto \eta'$
(the Iordansky force) is non-dissipative. While the parallel
to the vortex motion component 
of the friction force has a straight forward microphysical
interpretation in terms of scattering of the normal excitations off the 
vortex cores, the microscopic nature of the Iordansky force is 
controversial. We shall include (phenomenologically) this force
whenever the results are simple enough to disentangle the effect 
of a nonzero $\eta'$, otherwise we shall set $\eta' = 0$.
A nonzero $\beta''$ implies friction along the {\it average}
direction  of the vorticity, which is possible if
vortices are oscillating, or are subject to other
deformations in the plane orthogonal to the rotation.
It is reasonable to assume that for small perturbations the distortions
of the vortex lattice are small so that $\beta''\ll \beta, \beta'$.

\subsection{Perturbation equations for uniform rotations}

Virial equations of various order are constructed by taking the
moments of Eq.  (\ref{eq:euler}) with weights 1, $x_i$, $x_{i}x_{j}$ 
etc  and integrating over the volume $V_{\alpha}$ occupied by 
the fluid $\alpha$.  The first and second-order virial equations for 
trapped Bose-Einstein condesates at finite temperatures 
are derived in the Appendix A and the equations of second-order, which
determine the equilibrium state of the uniformly rotating 
condensate and thermal cloud, are obtained. Consider small perturbations 
from  equilibrium. The Eulerian variations of the
first-order virial equation [Appendix A, Eq. (A1)] 
lead to   
\baray
\label{eq:PERTURB1}
 \frac{d^2 V_{\alpha,i} }{dt^2}&=&
2\epsilon_{ilm}\Omega_m\frac{d}{dt} V_{\alpha,l}
+ (\Omega^2\delta_{ij}
- \Omega_i\Omega_j) V_{\alpha,j} 
 -\omega_i^2V_{\alpha,i}\nonumber\\
&-&{\cal S}_{\alpha\beta}\omega_S \beta_{ij}
\frac{d}{dt}\left( V_{S,j}-f V_{N,j}\right).
\earay
The variations of the second-order virial equation [Appendix A,
Eq. (A3)]  give
\baray
\label{eq:PERTURB2}
{d^2V_{\alpha,i;j}
\over dt^2}&=&2\epsilon_{ilm}\Omega_m
{dV_{\alpha,l;j}\over dt}
+(\Omega^2-\omega_i^2)V_{\alpha,ij}\nonumber\\
&-&\Omega_i\Omega_kV_{\alpha,kj}
-{\cal S}_{\alpha\beta}\omega_S\beta_{ik}\frac{d}{dt}
\left( V_{S,k;j}-fV_{N,k;j}\right)\nonumber\\
&+&\delta_{ij}\delta\Pi_{\alpha}
+\delta_{\alpha, N}\delta {\cal P}_{ij},
\earay
where $\delta{\cal P}_{ij}$ and $\delta \Pi_{\alpha}$ are the
Eulerian variation of the stress energy and pressure tensors defined
in the Appendix A, the 
virials $V_{\alpha, i}$, $V_{\alpha, i;j}$ are defined in terms of the
Lagrangian displacement $\xivec_{\alpha}$ as
\baray
\label{VALPHA1}
V_{\alpha, i}&=&\int_{V_{\alpha}}d^3 x\,\rho_{\alpha}\,
\xi_{\alpha,i},\\
\label{VALPHA2}
V_{{\alpha},i;j} &=& \int_{V_{\alpha}} d^3x \rho_{\alpha} \xi_{\alpha,i} x_j,
\earay
and $V_{{\alpha},ij} = V_{{\alpha},i;j}+ V_{{\alpha},j;i}$. 
Equations (\ref{eq:PERTURB1}) and (\ref{eq:PERTURB2}) assume the
density gradients are sufficiently smooth so that the position dependent
mutual friction tensor $\beta_{ij}$ can be approximated by a 
constant averaged value. 
Similarly, the factor $f =(\rho_S/\rho_N) (1-\epsilon)$ 
approximates   the position dependent ratio of the densities by
its average value over the volume. The volume reduction factor
$0\le \epsilon <1$  takes into account the fact that the integration is
over the overlap volume of the two fluids (since in general the 
fluids do not occupy the same volume in the 
background equilibrium).
The positive sign of $\epsilon$ corresponds to the case where the 
condensate is embedded in a thermal cloud of
a larger volume, as is frequently the case in experiments.

The center of mass and relative motions of the fluid  
can be decoupled by defining new linear combinations 
of the  virials
\baray\label{NEW_V1}
V_{ij\dots}&\equiv&  V_{S,ij\dots}+ V_{N,ij\dots},\\
\label{NEW_V2}
U_{ij\dots}&\equiv&  V_{S,ij\dots}-f V_{N,ij\dots}.
\earay
The decoupling is prefect when the hydrodynamic equations 
of the fluids (\ref{eq:euler}) are symmetric/antisymmetric with 
respect to an exchange of the indices labeling the fluids. This
is the case for the first-order virial equations (due to the surface 
boundary conditions). For the second-order virial equations the symmetry
is broken by either the difference in the pressure perturbations of
the condensate and the thermal cloud or by the viscosity which
acts only in the normal gas. 

At the first-order the center of mass motions of the two fluids are
trivial, since they can be eliminated by a linear transformation to
the reference frame where  $V_i =0$. The center of mass 
modes of second-order are governed by the equation
\be\label{eq:CM} 
\frac{d^2 V_{i;j} }{dt^2}=
2\epsilon_{ilm}\Omega_m\frac{d V_{l;j}}{dt}
+(\Omega^2-\omega_i)V_{ij}- \Omega_i\Omega_k V_{kj}
+\delta_{ij}\delta\Pi^{(+)},
\ee
which differs from Eq. (11) of Paper I by the form 
of the perturbation of the stress tensor
$\delta\Pi^{(+)} =\delta\Pi_N+\delta\Pi_S$.
Consequently, all the second harmonic modes  contained
in Eq. (\ref{eq:CM}) are identical to those derived in 
Paper I [see also Eqs. (\ref{int1})-(\ref{int4})]
except the quasi-radial pulsation modes, 
which require the explicit form of the pressure tensor
and couple the center of mass and relative motions (and  
therefore differ from those in Paper I).
The first and second-order virial equations for the relative
motions are
\baray
\frac{d^2 U_{i} }{dt^2}&=&
2\epsilon_{ilm}\Omega_m\frac{d U_{l}}{dt}
+(\Omega^2-\omega_i)U_i\nonumber\\
&-& \Omega_i\Omega_j U_{j}
-\omega_S \tilde\beta_{ij}\frac{d U_{j}}{dt},
\label{eq:relative_virial1}
\\			      
{d^2U_{i;j}\over dt^2}&=&2\epsilon_{ilm}\Omega_m
{dU_{l;j}\over dt}+(\Omega^2-\omega_i)U_{ij}\nonumber\\
&-&\Omega_i\Omega_k U_{kj}
+\delta_{ij}\delta\Pi^{(-)}
-2\Omega\tilde\beta_{ik}\frac{d}{dt}U_{k;j},
\label{eq:relative_virial2}
\earay
where we defined 
$			     
\tilde\beta_{ij} = [1+f]\beta_{ij}
$			     
and			       
$			     
\delta\Pi^{(-)}=\delta\Pi_S-f\delta\Pi_N. 
$
Note that the `effective' mutual friction tensor 
$\tilde\beta_{ij}$ could differ significantly from the 
`microscopic' one, $\beta_{ij}$, due to the re-scaling 
factor $f$ which depends on the volume mismatch 
and density fractions of the components that can be
manipulated experimentally.

\section{small-amplitude oscillations}
			    
We assume that the time dependence of Lagrangian displacements of the
condensate and the normal-fluid are of the form
\be\label{eq:XI_T}
\xivec_{\alpha}(x_i,t) = \xivec_{\alpha}(x_i)e^{i\sigma t}.
\ee			       
When these Lagrangian displacements are substituted in the definitions
of the virials, Eqs. (\ref{eq:relative_virial1}) and 
(\ref{eq:relative_virial2}) provide a set of characteristic equations
that determine the modes associated with the first and second-order
virial equations. 
We turn now to the characteristic equations that derive from 
the virial equations (\ref{eq:relative_virial1}) and 
(\ref{eq:relative_virial2}).

\subsection{First-order harmonic oscillations}	    

The characteristic equation for the first order relative
oscillations modes is
\baray%\label{eq:rel_1st_order}
\left(\sigma^2+\Omega^2-\omega_{i}^2\right) U_i + 
2i \epsilon_{ilm}\Omega_m\sigma U_l&&\nonumber\\
-\left(\Omega_i\Omega_j 
+2i \Omega\tilde\beta_{ij} \sigma \right)U_j=0.&&
\label{eq:monopole}
\earay
The components of Eq. (\ref{eq:monopole}) along and orthogonal 
to the rotation axis (i.e., those with odd and even-parity with 
respect to inversions of the $z$ direction) decouple and we find 
\baray
\sigma_{\pm}^{\rm odd}&=&\pm \sqrt{\omega_3^2-\tilde\beta^{''2}\Omega^2}
+i\tilde\bdp\Omega, 
\label{eq:mono_3}
\\
\sigma_{\pm}^{\rm eve} &=& 
\mp (1-\tilde\bp)\Omega +i\tilde\beta \Omega \pm \left(X^2+Y^2 \right)^{1/4}
\Omega\, e^{\mp i\psi/2}, \nonumber\\
\label{eq:mono_tor}
\earay
where $X=\omega_{\perp}^2/\Omega^2+\tilde\beta'^2-
\tilde\beta^2-2\tilde\beta'$,
$Y= {2\tilde\beta(1-\tilde\beta')}$  and ${\rm tan}~\psi =Y/X $.
The odd-parity modes are stable and damped; note that if 
$\tilde\beta^{''2}\Omega^2 \ge \omega_3^2$ the modes are purely 
imaginary. The asymptotics of the even-parity modes, neglecting the
rescaling of the $\beta$-coefficients due to the mismatch of the 
volumes of fluids, and for small $\zeta$ ({\it weak-coupling limit}) is
\baray\label{weak_asym}
{\rm Re}~ \sigma_{\pm} &\to & \mp \Omega (1+\zeta') 
\pm \Omega'+{\cal O}(\zeta^2) ,
\\
{\rm Im}~ \sigma_{\pm} &\to & \Omega  (1+\zeta') \zeta
-\Omega^2 (1+\zeta')^2\zeta/\Omega'+{\cal O}(\zeta^3),\nonumber\\
\earay
where $\Omega^{'2} = \omega_{\perp}^2-\Omega^2+\Omega^2
(1+\zeta')^2$; for large $\zeta$ ({\it strong-coupling limit}) we find
\baray\label{str_asym}
{\rm Re}~ \sigma_{\pm} &\to &
\pm \sqrt{\omega_{\perp}^2-\Omega^2}+{\cal O}(\zeta^{-2}),\\
{\rm Im}~ \sigma_{\pm}  &\to & \Omega (1+\zeta')/\zeta +
{\cal O}(\zeta^{-3}).
\earay
Equations. (\ref{weak_asym})-(\ref{str_asym}) are written at the zeroth
order of the expansion with respect to the small parameter 
$\delta = (\tilde\bp-\bp)/\bp$, which rescales the 
$\beta$ coefficients due to the mismatch of volumes of the fluids.
The ratios of the real to the  imaginary 
part of the modes increase linearly in both, the strong- and
weak-coupling limits for fixed $\zeta'$, hence the modes are 
weakly damped in both limits. An exception is the strong-coupling 
limit  when $\zeta'\propto \zeta$, in which case 
the ratio of the real to the  imaginary part tends to a constant 
value $\sqrt{\omega_{\perp}^2-\Omega^2}/\Omega$.

\begin{figure}[htb] % fig 1
\psfig{figure=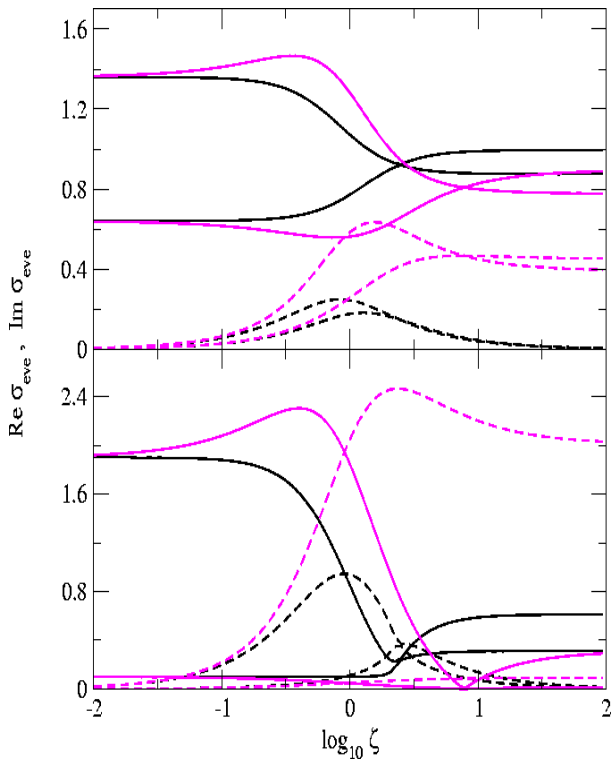,height=10cm,width=8.cm,angle=-90}
\caption{(Color online)
 The real ({\it solid lines}) and imaginary ({\it dashed lines}) parts 
 of the first order even-parity modes as functions of the 
 drag-to-lift ratio $\zeta$. The heavy and light
 lines correspond to $\zeta'= 0$ and $\zeta'=\zeta$ cases. 
 The upper and lower panels correspond to the rotation
 frequencies $\Omega/\omega_{\perp} 
= 0.36$ and $\Omega/\omega_{\perp} = 0.9$, respectively.
 The frequencies are in units of $\omega_{\perp}$. 
}
\label{MSfig:fig1}
\end{figure}

The real ({\it solid lines}) and imaginary  ({\it dashed lines}) parts
of the first order even-parity modes as functions of the drag-to-lift ratio
$\zeta$ are shown in  Fig. 1. 
The frequencies are in units of $\omega_{\perp}$.
The parameter values are those in the MIT experiment~\cite{MIT1}:  
$\omega_3^2/\omega_{\perp}^2 =0.057$ and we
assumed that the  overlap volume corresponding to  $\epsilon=
0.5$ and the condensate fraction is $20\%$ which implies $f= 1.125$.
The two panels in Fig. 1 are representative for the slow and fast rotation
limits with the values of the spin
frequency  $\Omega/\omega_{\perp} = 0.36$ (as is the case in the 
experiment~\cite{MIT1}) and $\Omega/\omega_{\perp} = 0.9$.
The heavy and medium lines correspond  to the cases $\zeta' = 0$ 
and $\zeta = \zeta'$, respectively. The generic features of the modes
(seen also for more complicated cases below) are  the asymptotically
constant values of the real and the vanishing of the imaginary 
parts, except for the case of $\zeta = \zeta'$ and strong-coupling
($\zeta \gg 1$) where the damping tends to a constant value asymptotically. 
The mode crossing occurs when the condition $\tilde\bp = 1$ is fulfilled.
This can be seen directly from the characteristic equations: the real
and imaginary parts degenerate to a single value since for $\tilde\bp
=1$ the first term on right hand side of  Eq. (\ref{eq:mono_tor})  
and  the phase of the second term in Eq. (\ref{eq:mono_tor}) vanish. 
The position of the mode crossing when $\zeta' = 0$ and  
$\zeta'=\zeta$ is different. The reason is that while the mode crossing 
is still defined by the condition  $\tilde\beta' =1$, the functional dependence 
of $\zeta$ on $\tilde \bp$  for $\zeta' = 0$ and $\zeta' = \zeta$ 
changes. The damping of the modes has two maxima, one at the position
of the mode crossing, the second at $\zeta = 1$.

In the weak-coupling limit $(\zeta \ll 1$) the splitting in the real 
parts of the modes is twice the rotation frequency consistent 
with Eq. (\ref{weak_asym}) and the centroid is at 
${\rm Re}~\sigma_{\rm eve} = \omega_{\perp}$.  
In the strong-coupling limit $(\zeta \gg 1$) the
splitting would have been absent if the fluids were to occupy the same 
volume (i.e., $\bp=\tilde\bp = 1$ when $\zeta\to \infty$). 
The strong-coupling expansion 
(\ref{str_asym}) to next-to-leading order in $\delta\ll 1$ reads
\be\label{str_asym_bis}
{\rm Re}~ \sigma_{\pm} \to \pm
\sqrt{\omega_{\perp}^2-\Omega^2}\pm \Omega\delta +{\cal O}(\zeta^{-2})
+{\cal O}(\delta^{2}).
\ee
Thus, the splitting of the real parts of the modes in the strong
coupling regime is a direct measure of the renormalization of
the $\beta'$ coefficient due to the mismatch of the volumes of the
condensate and the thermal cloud (provided the ratio $\rho_S/\rho_N$ is
known). The centroid of the splitting is
defined by the first term in Eq. (\ref{str_asym_bis}). 
The overall behavior of the modes in the fast  
(or near critical, $\Omega/\omega_{\perp} \le 1$) 
rotation regime, shown in the lower panel of Fig. 1, is qualitatively
the same as for the slow rotations. The splittings of 
the real parts of the modes increases in the strong and weak
coupling limits linearly with $\Omega$. In the fast
rotation regime, the modes are either partially (for $\zeta'=0$) or
fully  (for $\zeta'=\zeta$) damped in the strong-coupling limit.  Note
that the point where the modes cross is independent of
$\Omega$.

\subsection{second-order harmonic oscillations}

This section contains a formal derivation of the
characteristic equations for $l=2$  modes in the inviscid and 
viscose normal-fluid cases. The numerical results and the 
physical properties of these modes are discussed in the 
following section.

\subsubsection{Inviscid limit}

It is instructive to consider the second-order harmonic small-amplitude 
oscillations first in the inviscid limit ($\nu\to 0$).
The characteristic equations for the second-order center of mass
and relative oscillations modes are 
\baray
\label{eq:CM_virial_2}
\sigma^2V_{i;j}+(\Omega^2-\omega_i^2) V_{ij}
+2i\epsilon_{ilm}\Omega_m\sigma V_{l;j}&&\nonumber\\
-\Omega_i\Omega_k V_{kj}+\delta_{ij}\delta\Pi^{(+)}&=&0,\\
\sigma^2 U_{i;j}+(\Omega^2-\omega_i^2) U_{ij}
+2i\epsilon_{ilm}\Omega_m\sigma U_{l;j}&&\nonumber\\
-2i\Omega\tilde\beta_{ik}\sigma
U_{k;j}-\Omega_i\Omega_k U_{kj}+\delta_{ij}\delta\Pi^{(-)}&=&0,
\label{eq:relative_virial_2}
\earay
and the even and odd-parity modes can be treated separately. 
Equation (\ref{eq:CM_virial_2}) for center of mass oscillations differs
from the analogous Eq. (11) of Paper I by the variation
of the pressure tensor $\delta_{ij}\delta\Pi^{(+)}$; 
since the transverse-shear ($l=2$ and $\vert m\vert
= 1$) and the toroidal ($l=2$ and $\vert m\vert = 2$) modes do not involve 
this quantity, these modes are identical to those derived earlier, 
see Eqs. (\ref{int1}) and (\ref{int4}). 
The quasi-radial pulsation modes couple the center of mass and 
relative oscillations and will be examined below.

We start with the {\it toroidal modes} of relative oscillations, which
are governed by the equations
\baray
\label{tor1}
\left[\sigma^2-2i\Omega\tilde\beta\sigma-2(\omega_{\perp}^2-\Omega^2)\right]
(U_{11}-U_{22})&&
\nonumber\\+4i\Omega(1-\tilde\beta')\sigma U_{12}&=&0,
\\
\label{tor2}
\left[\sigma^2-2i\Omega\tilde\beta\sigma-2(\omega_{\perp}^2-\Omega^2)\right]
U_{12}&&
\nonumber\\-i\Omega(1-\tilde\beta')\sigma (U_{11}-U_{22})&=&0 .
\earay
The associated characteristic equation and the solutions are 
identical to  Eq. (\ref{eq:mono_3}) and (\ref{eq:mono_tor}), respectively,
if we replace in these equations the radial trapping frequency  by
 $\omega_{\perp}^{*2}= 2\omega_{\perp}^2-\Omega^2$; 
clearly, with the substitution above,
the limiting behavior of the modes for large and small frictions
are the same as for the first order harmonic modes. 
The {\it transverse-shear} (odd-parity) modes are given by the components of
Eq. (\ref{eq:relative_virial_2}) for the tensors 
$U_{i;3}$ and $U_{i3}$, $i = 1,2$:
\baray
\label{tra1}
(\sigma^2-2i\Omega\tilde\beta\sigma)U_{1;3}
+2i\Omega(1-\tilde\bp)\sigma U_{2;3}&&
\nonumber\\
-(\omega_{\perp}^2-\Omega^2)U_{13} &=& 0,\nonumber\\
\\
\label{tra3}
(\sigma^2-2i\Omega\tilde\bdp\sigma) U_{1;3}
+(\sigma^2-2i\Omega\tilde\bdp\sigma-\omega_3^2) U_{13} &=&0,
\nonumber\\
\earay
and the remaining two equations are obtained from Eqs. (\ref{tra1}) and
(\ref{tra3}) via the $1\leftrightarrow 2$ interchange.
The corresponding characteristic equation
is of order 7.

By a suitable combination of the even-parity
components of Eq. (\ref{eq:CM_virial_2}) we find the following set 
of equations for the virial combinations $V_{11}+V_{22}$,
$V_{1;1}-V_{2;2}$, and $V_{33}$  describing the {\it pulsation modes}
\baray 
\left(\sigma^2/2-\omega_{\perp}^2+\Omega^2\right)
+\left(V_{11}+V_{22}\right)&&\nonumber\\
-2i\Omega\sigma (V_{1;2}-V_{2;1})
-\left(\sigma^2- 2\omega_3^2\right) V_{33} &=&0,\\
\sigma^2 (V_{1;2}-V_{2;1}) +i\Omega\sigma (V_{11}+V_{22})
&=&0,
\\
(\sigma^2-2\omega_3^2) V_{33}+2\delta\Pi^{(+)}&=&0,
\earay
where the tensor $\delta\Pi^{(+)}$ is defined by Eq. (B10) of the
Appendix B. It involves 
the same set of virials and, in addition, couples these equations to the 
virials describing relative motions.
Analogous manipulations on Eq. (\ref{eq:relative_virial_2}) lead to 
\begin{widetext}
\baray
\left(\sigma^2/2-i\Omega\tilde\beta\sigma
-\omega_{\perp}^2+\Omega^2\right)(U_{11}+U_{22})
-\left(\sigma^2- 2i \Omega\tilde\bdp \sigma - 2 \omega_3^2 \right)
U_{33}
- 2i\Omega(1-\tilde\bp)\sigma(U_{1;2}-U_{2;1})&=&0,
\\
(\sigma^2-2 i\Omega\tilde\beta\sigma)(U_{1;2}-U_{2;1})
+i\Omega(1-\tilde\bp)\sigma (U_{11}+U_{22}) &=& 0,
\\
(\sigma^2-4i\Omega\tilde\bdp-2\omega_3^2) U_{33}+2\delta\Pi^{(-)}&=&0,
\earay
with   $\delta\Pi^{(-)}$ defined by Eq. (B11) of the
Appendix B. The resulting characteristic equations 
for the quasiradial pulsation modes ($l = 2$, $m = 0$) is of 
order 9. 

\subsubsection{Including viscosity of thermal cloud}

The kinematic viscosity of the normal-fluid can be included in the 
tensor virial formalism perturbatively and the Eulerian variations 
of the viscose stress tensor are computed in the Appendix B. 
The virial equation that  contain the second-order harmonic $l=2$
and $-2\le m \le 2$ modes in the presence of normal cloud 
viscosity are obtained upon substituting Eq. (B3) 
of the Appendix B in Eqs. (\ref{eq:CM_virial_2}) and  
 (\ref{eq:relative_virial_2}):
\baray
\sigma^2 V_{i;j}+2i\epsilon_{ilm}\Omega_m \sigma V_{l;j}
+(\Omega^2-\omega_i^2)V_{ij}-\Omega_i\Omega_k V_{kj}
+\delta_{ij}\delta\Pi^{(+)}&&
\nonumber\\
- \frac{5i\nu \sigma}{1+f}  \left(\frac{V_{i;j}}{\bar a_j^2}
+\frac{V_{j;i}}{\bar a^2_i}-\frac{V_{ll}}{3\bar a_l^2}\delta_{ij}\right)
+ \frac{5i\nu \sigma}{1+f} \left(\frac{U_{i;j}}{\bar a_j^2}
+\frac{U_{j;i}}{\bar a^2_i}-\frac{U_{ll}}{3\bar
a_l^2}\delta_{ij}\right)
&=& 0,
\label{eq:CM_virial_3}
\\
\sigma^2 U_{i;j}+2i\epsilon_{ilm}\Omega_m
\sigma U_{l;j}+(\Omega^2-\omega_i^2)U_{ij}-\Omega_i\Omega U_{kj}
+\delta_{ij}{\delta\Pi}^{(-)}
-2i\Omega\tilde\beta_{ik}\sigma U_{k;j}&&
\nonumber\\
+\frac{5if\nu \sigma}{1+f} \left(\frac{V_{i;j}}{\bar a_j^2}
+\frac{V_{j;i}}{\bar a^2_i}-\frac{V_{ll}}{3\bar a_l^2}\delta_{ij}\right)
- \frac{5if\nu \sigma}{1+f} \left(\frac{U_{i;j}}{\bar a_j^2}
+\frac{U_{j;i}}{\bar a^2_i}-\frac{U_{ll}}{3\bar
a_l^2}\delta_{ij}\right) &=& 0.
\label{eq:relative_virial_3}
\earay
Note that Eqs. (\ref{eq:CM_virial_3}) and (\ref{eq:relative_virial_3})
are symmetric with respect to the simultaneous interchanges
$U\leftrightarrow V$, $f \nu \leftrightarrow \nu$ and 
$\delta\Pi^{(-)}\leftrightarrow\delta\Pi^{(+)}$, however
Eq. (\ref{eq:relative_virial_3}) contains a term 
$\propto \beta_{ik}$ which is absent
in Eq. (\ref{eq:CM_virial_3}). To keep the presentation compact we
will quote below only the components of
Eq. (\ref{eq:relative_virial_3}); the analogous
components of Eq. (\ref{eq:CM_virial_3}) follow by 
the interchanges above and by setting $\beta_{ij}  = 0$
(this will be abbreviated as the $UV$-transformation).

The {\it transverse-shear modes} are determined by the eight components 
of the Eqs. (\ref{eq:CM_virial_3}) and (\ref{eq:relative_virial_3})
which are  odd in index 3, i.e., $V_{3;i}$,  $V_{i;3}$,  $U_{3;i}$
and  $U_{i;3}$, where $i=1,2$. The odd equations for the virials 
describing the relative motions are
\baray  
\label{axisym_modes:v1.5}
\left(\sigma^2 -2 i \Omega\tilde\beta\sigma +i \chi
f \nu\sigma\right) U_{1;3} 
-\left(\omega_{\perp}^2-\Omega^2+if  \nu\sigma\right)U_{13} 
+2i\Omega(1-\tilde\bp)\sigma U_{2;3}  
-i\chi f \nu\sigma V_{1;3}+if  \nu\sigma V_{13}&=&0,
\\
\label{axisym_modes:v1.7}
\left(\sigma^2 -2 i \Omega\tilde\bdp\sigma-\omega_3^2 
- i f \nu\sigma\right) U_{13}
+\left(\sigma^2 - 2\Omega i\tilde\bdp\sigma  
- i \chi f \nu\sigma\right)U_{1;3} 
+if \nu\sigma V_{13}-i\chi f \nu\sigma V_{1;3} &=& 0,
\earay
two additional equations are obtained from Eqs.
(\ref{axisym_modes:v1.5}) and (\ref{axisym_modes:v1.7}) 
via the $1\leftrightarrow 2$ interchange
and the analogous equations for the center of mass virials
are obtained via the $UV$ transformation.

Here $\chi\equiv 1 -\bar a_1^2/\bar a_3^2$ 
and $\nu' = 5\nu/\bar a_1^2(1+f)$ and 
the primes have been dropped in the equations. (Note that
we used the relation  $U_{ij} = U_{i;j} + U_{j;i}$ to 
manipulate the components of Eq. (\ref{eq:relative_virial_3})   
to the form above.) 
The resulting two sets of equations for the center of mass and 
the relative virials decouple 
in the  limit $\nu \to 0$, as they should.
The dissipation in the  first set is 
driven by the viscosity of the normal-fluid. 
In the second set the normal-fluid
viscosity contributes to the damping of the
Lagrangian displacements which are 
orthogonal to those damped by the mutual
friction, and in addition,  the mutual friction 
damping time scale is renormalized form $2\Omega\tilde\beta$ to 
$2\Omega\tilde\beta+(1-\chi)f   \nu$;
(this renormalization vanishes for a sphere, since then $\chi=1$).
The resulting characteristic equation is of order 12.

The {\it toroidal modes} are determined by the even in index 3 components
of Eqs. (\ref{eq:relative_virial_2}) and (\ref{eq:CM_virial_2})
written for the virials  $V_{i;i}$,  $V_{i;j}$,  $U_{i;i}$ and  $U_{i;j}$, 
where $i,j=1,2$. These equations can be manipulated to a 
set of four equations and those 
for the relative virials are
\baray
\left[\sigma^2-2 i \Omega\tilde\beta\sigma
-2i f \nu\sigma-2(\omega_{\perp}^2-\Omega^2)\right]
(U_{11}-U_{22})+4i\Omega(1-\tilde\beta')\sigma U_{12}
+2if \nu\sigma (V_{11}-V_{22})&=&0,
\\
\left[\sigma^2-2 i \Omega\tilde\beta\sigma
-2i f \nu\sigma-2(\omega_{\perp}^2-\Omega^2)\right]
U_{12}-\Omega(1-\tilde\beta')i\sigma (U_{11}-U_{22}) 
+2if \nu \sigma V_{12}&=&0 ,
\earay
the analogous equations for the center of mass virials follow 
by the $UV$ transformation.
As above, the effect of the kinematic viscosity is the coupling of the
center of mass and relative modes and the renormalization 
of the damping due to the mutual friction
$\Omega\tilde\beta\to\Omega\tilde\beta+ \nu$.
The corresponding characteristic equations is of eighth order.

The {\it pulsation modes} are determined by the equations
which are even in index 3. For incompressible ellipsoids these should be
supplemented by the conditions of vanishing of the divergence of perturbations
of each fluid [Paper I, Eq. (38) and its analog for the $U_{ii}$
virials]. We shall proceed directly to the compressible fluid case. 
Combining  the equations for the virials  
$V_{i;i}$,  $V_{i;j}$,  $U_{i;i}$  and  $U_{i;j}$, 
where $i=1,2,3$ and $j=1,2$, we find two (coupled) subsets of
equations, the first involving perturbations which are orthogonal to
the spin axis 
\baray 
(\sigma^2 - 2i \Omega \tilde\beta\sigma) (U_{1;2}-U_{2;1}) 
+i\Omega(1-\tilde\bp)\sigma (U_{11}+U_{22})&=&0,
\earay
and the second subset which couples the orthogonal to the spin parallel 
perturbations [$\tilde\nu \equiv (1-\chi)\nu]$
\baray
\left(\frac{\sigma^2}{2}-if  
\nu\sigma-i\Omega\tilde\beta\sigma-\omega_{\perp}^2+\Omega^2 
\right)(U_{11}+U_{22})- 2i\Omega(1-\tilde\bp)
\sigma (U_{1;2}-U_{2;1})\nonumber\\ 
-\left(\sigma^2- 2i \Omega\tilde\bdp\sigma 
- 2 \omega_3^2 - 2i f\tilde \nu\sigma\right) U_{33}
+if \nu\sigma (V_{11}+V_{22})  +2if \tilde\nu \sigma V_{33} 
&=&0,\\
\left(\frac{\sigma^2}{2}-\frac{2i}{3} f \tilde\nu\sigma
-i\Omega\bdp\sigma-\omega_3^2\right)U_{33}
+\frac{i}{3} f \nu\sigma (U_{11}+U_{22})+\delta\Pi^{(-)}
+ \frac{2i}{3}f \tilde\nu\sigma V_{33} - \frac{i}{3}f \nu\sigma
(V_{11}+V_{22}) &=& 0,
\earay 
Note that the kinematic viscosity renormalizes the damping 
along the spin axis as $\Omega\tilde\bdp\to \Omega\tilde\bdp+
 (1-\chi)\nu$ and in the plane orthogonal to the spin axis as 
$\Omega\tilde\beta\to \Omega\tilde\beta+ \nu$. Unlike the 
$m\neq 0$ modes, the center of mass and relative 
motions remain coupled in the limit $\nu\to 0$ 
due to the differences in the pressure perturbations 
of the superfluid and normal-fluid. The characteristic equation
for the pulsation modes is of order 9. 
\end{widetext}

\subsection{Numerical solutions}

The fluid oscillations can be characterized in terms of 
Lagrangian displacement fields. For a single fluid,
the $l=2$ displacement fields are  shown schematically in Fig.  2
for (top-to-bottom)  $m=1$,  $m=2$, and $m=0$ modes. 
The $m=\pm 1$ modes are antisymmetric
with respect to the body rotation plane and involve
relative shearing of the ``northern'' and ``southern hemisphere''. 
The $m= \pm 2$ modes are non-axisymmetric, i.e., they tend to break the
axial symmetry of the body by deforming them into triaxial form. The 
$m= 0$ modes are symmetric with respect to the body
rotation plane and are the generalization of the 
ordinary pulsations of a sphere to the rotation.

\begin{figure}[htb] % fig 2
\psfig{figure=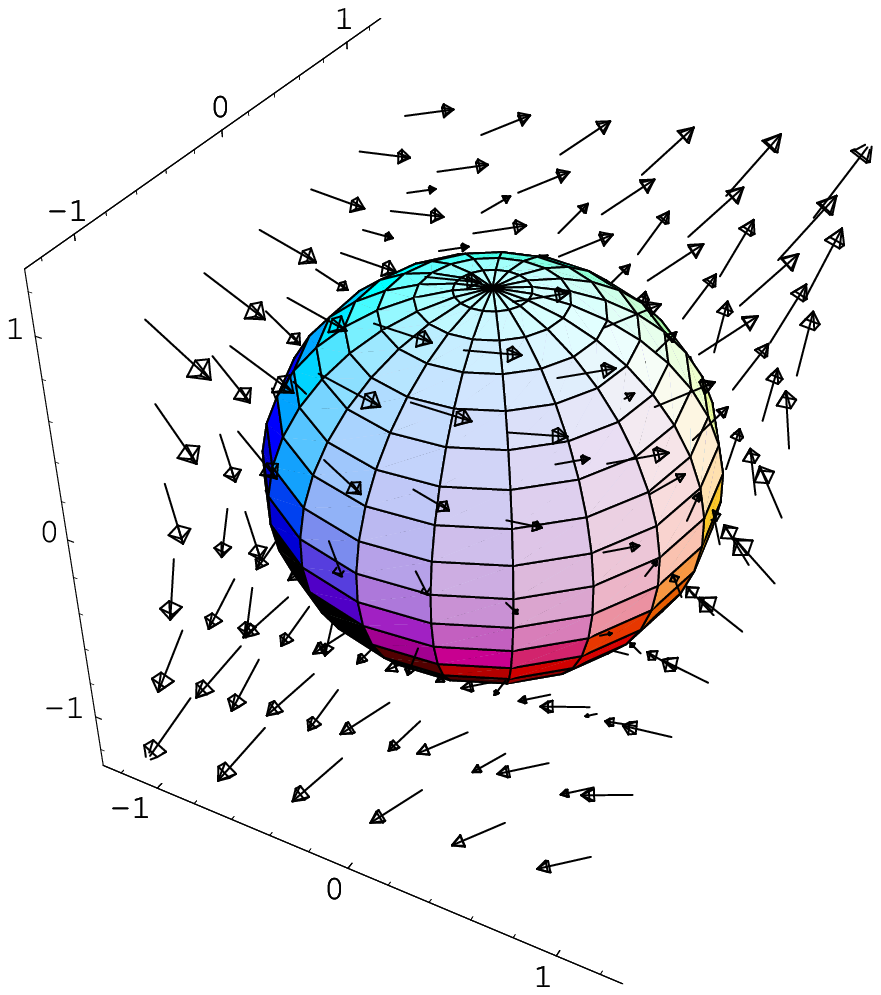,height=5cm,width=6cm,angle=0.0}
\psfig{figure=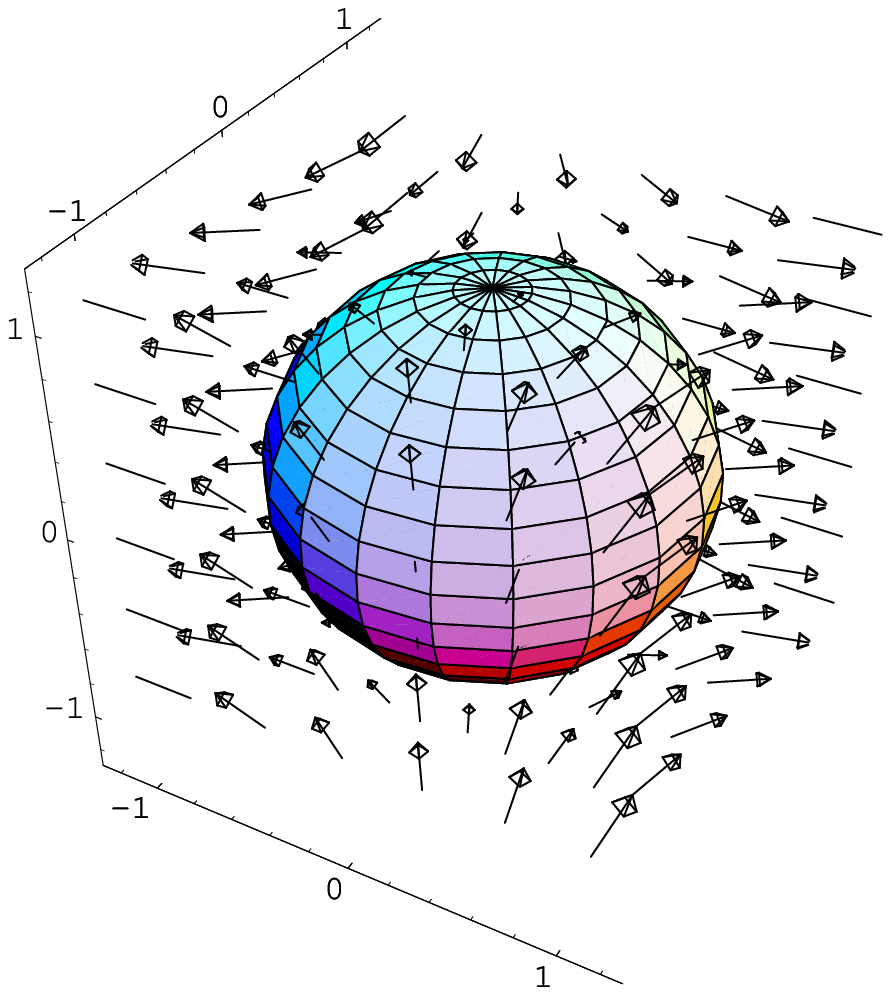,height=5cm,width=6cm,angle=0.0}
\psfig{figure=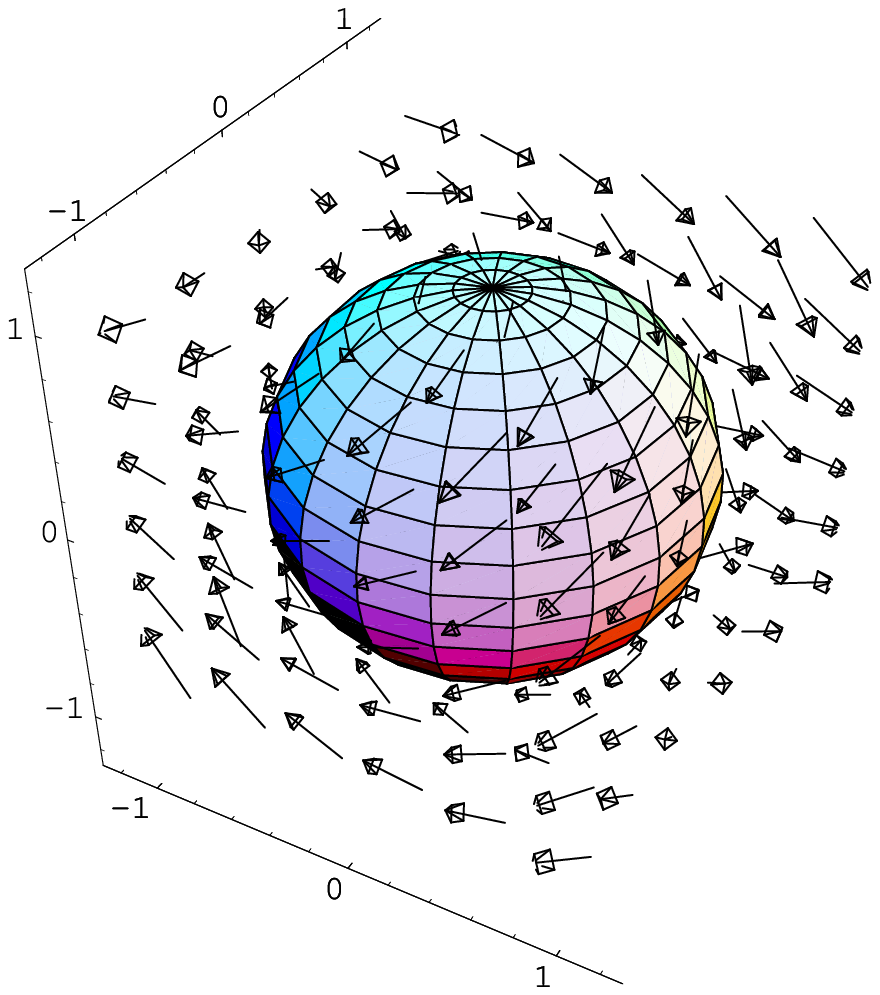,height=5cm,width=6cm,angle=0.0}
\caption{(Color online)
Schematic illustration of the displacement fields for 
$l=2$ and (top-to-bottom) $m =  1$,  $m = 2 $, and 
$m = 0$ oscillation modes. The rotation vector is along the vertical
$z$ axis and is directed from the ``south'' to the ``north pole''.
}
\label{MSfig:fig2}
\end{figure} 

In the two fluid setting, 
the superfluid and normal-fluid motions can be visualized 
in terms of either the two ``proper''displacements $\xivec_{S,N}$ 
[defining the $V_{S,N}$ virials, Eq. (\ref{VALPHA2})] or 
by their linear superpositions $\xivec_+= (\xivec_N+\xivec_S)$,
$\xivec_- =( \xivec_--f\xivec_N)$ [defining the 
$V$ and $U$ virials, Eqs. (\ref{NEW_V1}) and 
(\ref{NEW_V2})]. Note that the 
displacement fields and, hence, their linear superpositions, 
are linear in coordinates and the vector fields in Fig. 2 
can be associated with either displacement above; (of course,
the coefficients in the generic relation $\xi_i = L_{ij}x_j$ are 
different in each case). 
The relative oscillations can be viewed as a superposition of 
two antiparallel displacement fields  $\xivec_{S,N}$
each following the patterns shown in Fig. 2 and, thus, involving 
countermotions of the two fluids. 
Similarly, the center of mass modes are a superposition of the 
parallel displacement fields of the fluids and correspond
to a comotion of the fluids.

We now turn to the numerical solutions of the 
characteristic equations for the second-order virials. 
The results shown below were obtained for the parameter values 
$\omega_3^2/\omega_{\perp}^2 =0.057$, $f = 1.125$. (The latter
corresponds to the overlap volume $\epsilon=0.5$ and the condensate 
fraction  $20\%$)

\begin{figure}[htb] % fig 3
\centerline{\psfig{figure=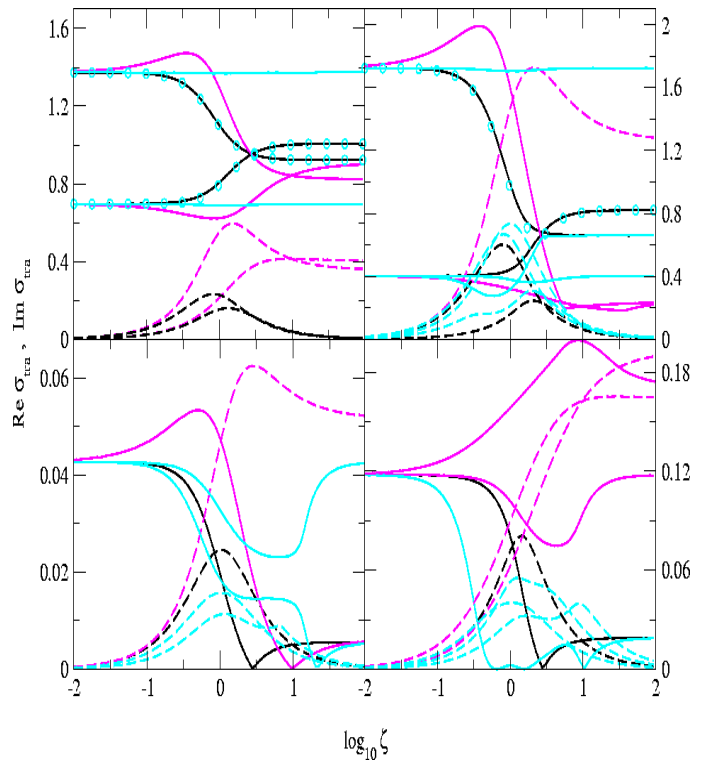,height=10cm,width=9cm,angle=-90}}
\caption{(Color online)
The dependence of the second-order transverse-shear 
($l=2$  $m = \pm 1$) modes on the drag-to-lift ratio $\zeta$;
the real and imaginary parts are shown by the solid and dashed lines.
The heavy and the medium lines correspond to 
the cases $\zeta'= 0$ and $\zeta'=\zeta$ with $\nu = 0$. The light
lines show the modes for $\nu = 2 \Omega \beta$ and  $\zeta'= 0$. 
The rotation frequency is   $\Omega/\omega_{\perp} = 0.36$  
in  the left panels and  $\Omega/\omega_{\perp} = 0.72$ in the  
right panels; the top and bottom panels 
show the high and the low frequency modes, respectively. The circles 
show the viscose modes that coincide with the modes in the inviscid 
limit.
}
\label{MSfig:fig3}
\end{figure} 
Fig. 3 shows the real ({\it solid lines}) and imaginary 
({\it dashed lines}) parts of the second-order transverse-shear 
modes ($l = 2, m = \pm 1$).
The heavy lines correspond to the inviscid case  with $\zeta'=0$; 
the medium lines correspond to the inviscid case but 
with   $\zeta'=\zeta$,  and the light 
lines - to the viscous  normal-fluid case with   
$\zeta'=0$. The coefficient $\beta''$ is set equal to zero everywhere.
The left and right panels correspond to the  rotation 
frequencies $\Omega/\omega_{\perp} = 0.36$ and $\Omega/\omega_{\perp} = 0.72$.
The characteristic equation for the transverse-shear modes is of order
7, however the solutions appear as complex-conjugate pairs, 
i.e., there are only three distinct solutions plus a zero-frequency
mode; this mode degeneracy reflects the axial symmetry of the
problem. An interesting feature seen in the inviscid limit, e.g., 
when $\zeta'=0$, is the mode crossing at $\tilde\bp = 1$, where
the real and imaginary parts of the two high-frequency 
modes coincide and the real part of the 
third low-frequency mode vanishes. 
The mode crossing occurs because the 
terms  $\propto 1-\tilde\bp$  in Eq. (\ref{tra1}) and its analog 
equation under the $1\leftrightarrow 2$ replacement
vanish, i.e., these equations become 
invariant under the exchange of the indices 1 and 2. 
Nonzero $\zeta'$  causes a  shift in the value of 
$\zeta$ where the modes cross,  which is a simple consequence 
of the different functional dependence of $\zeta$ on $\tilde \bp$  
for the cases $\zeta' = 0$ and $\zeta' = \zeta$. Note that the crossover from
the weak- to the strong-coupling regime is monotonic for the case 
$\zeta'=0$ while the functions acquire maximum or 
minimum at intermediate values of $\zeta$ when $\zeta'\neq 0$.
%%%%%%%%%%%%%%%%%%%%%%%%%%%%%%%% asymptotics 
The asymptotics of the real and imaginary parts of the
transverse-shear modes are the same as for the first-order 
even-parity modes (Fig. 1): the real parts of the  
modes tend asymptotically to 
constant values and the imaginary parts  vanish except for the 
case $\zeta = \zeta'$ and strong-coupling ($\zeta \gg 1$) where 
the damping tends to a constant value asymptotically. The splitting
of the modes is twice the rotation frequency in the weak-coupling
regime and is proportional to $f\Omega$ in the strong-coupling
regime, where $f$, defined via the relation $\tilde \beta' =
(1+f)\beta'$, reflects the mismatch in the volume of the
normal-fluid and superfluid.
The damping of the modes is negligible in the 
limits of both large and small $\zeta$ (except $\zeta'=\zeta$ and
$\zeta\to \infty$ limit) and is
maximal for $\zeta\sim 1$. This feature is generic to the damping of 
all modes and is a result of low vortex mobility for 
motions away or towards the spin axis: in the small $\zeta$ limit the
lattice moves with the  superfluid; in the large  $\zeta$ limit 
it is locked to the normal-fluid.
%%%%%%%%%%%%%%%%%%%%%%%%%%%%%%% fast rotations 
Comparing the modes for the slow and fast rotation cases (left and
right panels of Fig. 3) reveals the following differences: 
(i) the splittings of the real parts of the modes are 
larger in the strong- and weak-coupling limits 
consistent with the linear scalings with $\Omega$;
(ii) the two high-frequency modes are nonoscillatory in the 
crossover region $\zeta\sim 1$.  

%%%%%%%%%%%%%%%%%%%%%%%%%%%%%%% viscosity
When the viscosity of the normal-fluid is taken into account,
the number of modes double, 
since the relative and center of mass modes mix. 
The solutions of the 12th order characteristic equation for the
transverse-shear modes in the case of viscous thermal cloud are shown
in Fig. 3 by  light lines. The value of the kinematic
viscosity is fixed at $\nu = 2\beta\Omega$. (We note here that the
parameter $\nu/\beta\Omega$ is not small in our examples below, and
the perturbation expansion for the relative modes breaks down for 
$\nu/\beta\Omega\sim 1$. The quoted value was chosen to make the
differences between the inviscid and viscous case visible on the
scale of the figure.)  The solutions to the secular 
equations for the modes appear as complex-conjugates
(reflecting the axial symmetry of the problem), i.e.,  there are six 
distinct solutions for the transverse-shear modes.  
Consistent with the perturbative treatment of the viscosity effects
a subset of modes in Fig. 3 shows only small deviations from the
inviscid case, mostly in the crossover region 
between the strong and weak-couplings ($\zeta\sim 1$). 
The remainder modes are almost independent of the 
mutual friction and  are associated with the center of mass
oscillations modes. The generic features of the relative oscillation
modes are unchanged when viscosity is included  (the mode-crossing, 
asymptotic splittings of the real parts and the asymptotic decay of 
the imaginary parts). The shift in the position of the 
crossing point compared to the  inviscid and 
$\zeta'=0$ case  is due to the rescaling of the
$\tilde\beta'$ parameter by the terms $\propto  \chi\nu$. Note that
the damping of the center of mass modes is solely due to the viscosity
and the bell-shaped imaginary parts seen in Fig. 2 are the
consequence of our assumption that $\nu\propto \beta$ (assigning a
constant value to the kinematic viscosity would not change the
qualitative picture above, but would render the damping of the
center of mass oscillations independent of $\zeta$).

\begin{figure}[htb] % fig 4
\centerline{\psfig{figure=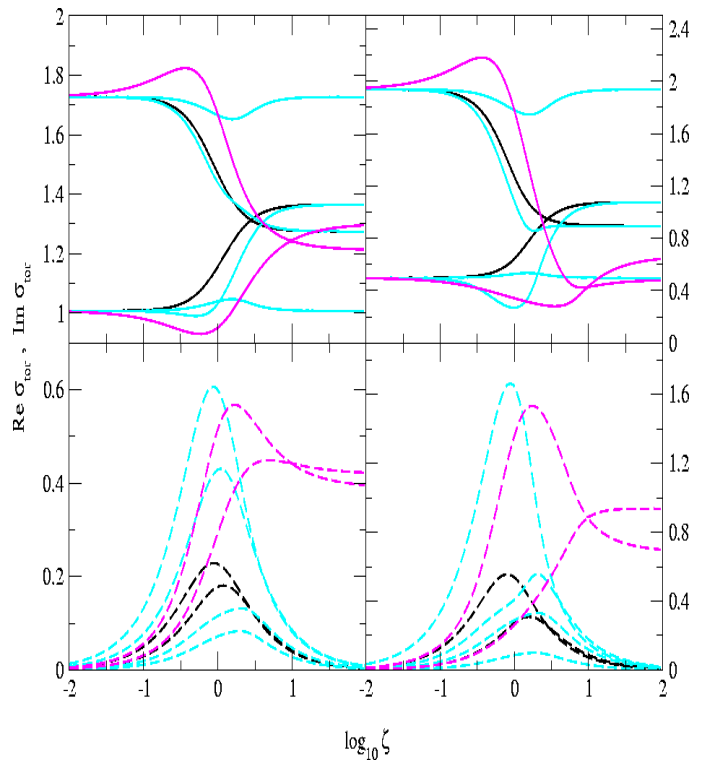,height=10cm,width=9cm,angle=-90}}
\caption{(Color online)
The dependence of the second-order toroidal ($l=2$  $m = \pm  2$) 
modes on the drag-to-lift ratio $\zeta$. The left and right panels
correspond to 
$\Omega/\omega_{\perp} = 0.36$ 
and  $\Omega/\omega_{\perp} = 0.72$. The labeling 
conventions are the same as in Fig. 3. }
\label{MSfig:fig4}
\end{figure} 

The solutions of the fourth-order characteristic equation for the 
toroidal modes ($l = 2, m = 0$) are displayed in
Fig. 4 (conventions are the same as in Fig. 3). Only two distinct
solutions exist in the inviscid limit, since the modes appear as 
complex-conjugates due to the axial symmetry.
Since the main properties of the toroidal modes are similar to those
of the transverse-shear modes we only briefly list
these features: (i) the pair of solutions degenerate to a single value
(mode-crossing) for $\tilde\beta'=1$ in the $\zeta'=0=\nu$
limits, since then Eqs. (\ref{tor1}) and (\ref{tor2}) become
identical. (ii)  For  $\zeta'=\zeta$ the value of $\zeta$ at which the crossing
occurs shifts to larger values. (iii) The real parts of the modes are
asymptotically constant; the two imaginary parts decay asymptotically to
zero and are maximal at ${\rm log}_{10} 
\zeta =0$ and at the mode-crossing point,
respectively. (iv) For $\zeta =\zeta'$ the imaginary parts are constants 
in the limit $\zeta\to \infty$. The asymptotic values of the
frequencies of the toroidal modes are numerically larger that 
those of the transverse-shear modes. 
In the viscose case  the mixing of the modes doubles the 
number of distinct toroidal modes and the additional modes can be
attributed to the center of mass oscillations. The deviation 
of the modes from the inviscid case are small in Fig. 4 and the perturbative 
treatment of the viscous effects is justified.

\begin{figure}[htb] % fig 5
\centerline{\psfig{figure=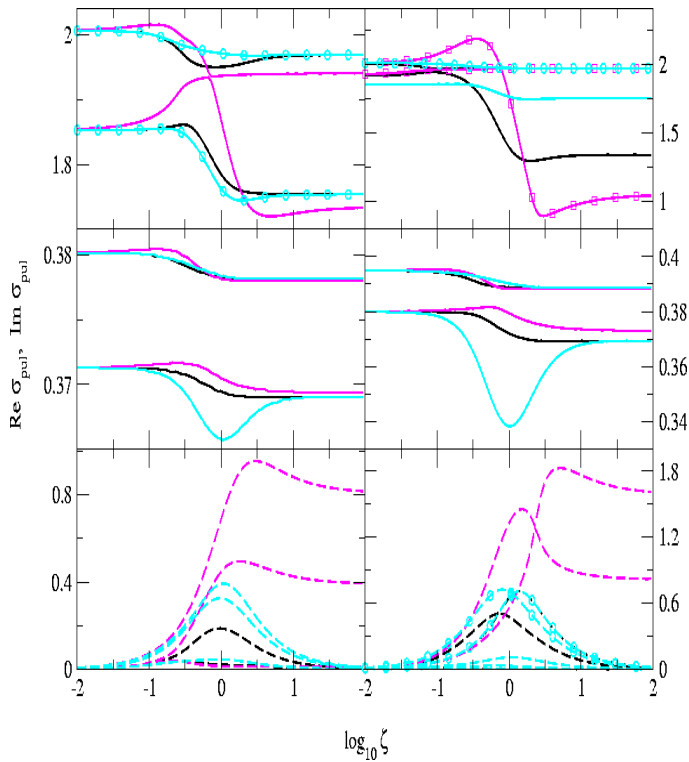,height=10cm,width=9cm,angle=-90}}
\caption{(Color online)
The dependence of the second-order pulsation 
($l=2$  $m = \pm 0$) modes on the drag-to-lift ratio 
$\zeta$. The left and right panels
correspond to $\Omega/\omega_{\perp} = 0.36$ 
and  $\Omega/\omega_{\perp} = 0.72$. The labeling 
conventions are the same as in Fig. 3 (the circles and squares show 
the modes in the viscose case which coincide with those in the
inviscid limit).
}
\label{MSfig:fig5}
\end{figure} 

Figure  5 displays the real and imaginary parts of the pulsation ($l=2$
$m=0$) modes. The distinctive feature of the pulsation modes is that
the center of mass and relative oscillations mix even in the inviscid
limit because  pressure perturbations for the 
normal-fluid and the condensate are governed by different adiabatic
indices.  The ninth-order secular equation 
for the pulsation modes has four distinct solutions plus a
zero-frequency mode (as before, because of the axial symmetry the
modes appear as complex-conjugates). Unlike the modes with $m\neq 0$
there is no generic crossing point for the pulsation modes.  
Including the viscosity of the thermal cloud causes small modifications 
in the oscillations modes (but does not change their overall number). 
The modes are mostly affected
in the crossover regime $\zeta\sim 1$ because of our assumption
$\nu \propto \beta$. The asymptotic behavour of the damping
of the modes is the same as for the transverse-shear and
toroidal modes.

\section{Conclusions and summary}

The analysis above shows  that the finite temperature
oscillation modes of a harmonically confined rotating Bose-Einstein 
gas contain rich information on the physics of condensate, the 
thermal cloud and their interactions via the vortex lattice.
The features of the oscillation modes are summarized below.

The modes separate into two classes
corresponding to the center of mass and relative 
oscillations of the thermal cloud and the condensate.
The first class of modes involves oscillations of the combined
mass of the cloud and the condensate, and is the analog of the
first sound in nonrotating superfluids.  The second class 
of modes carries zero net mass-current, involves  entropy 
oscillations of the thermal cloud, and hence, is the analog 
of the second sound.

The two classes of modes are independent when the normal cloud 
is inviscid. An 
exception are the radial pulsations, which are coupled due 
to the difference in adiabatic pressure perturbations of the cloud and the 
condensate. If the thermal cloud is viscous, the center of mass
and relative oscillations are coupled for all types of modes.

Two distinct mechanisms of damping of perturbations from the
equilibrium rotating state are operative: the mutual friction between 
the condensate and the thermal cloud, mediated by the vortex 
lattice, damps the relative  oscillations. The
kinematic viscosity damps the center of mass oscillations
of the thermal cloud. When the two classes of modes are
coupled, the kinematic viscosity renormalizes the damping
of the relative oscillation modes.

{\it First order harmonic oscillations.} At finite temperatures 
the lowest-order non-trivial (i.e., different from ordinary 
rotation) modes of oscillations belong to the first order 
($l=1$) harmonics. These modes represent relative oscillations 
of the thermal cloud and are described by  our Eqs. (\ref{eq:mono_3}) 
and  (\ref{eq:mono_tor}).
According to  Eq. (\ref{eq:mono_3}) the eigenmodes of 
the odd-parity oscillations should coincide with $\omega_z$ with zero
damping unless there is nonzero friction orthogonal to the equatorial
plane (which would indicate convective motions or vortex bending).    
The even-parity modes are described by Eq.  (\ref{eq:mono_tor})
and they carry information on a number of  
macroscopic and microscopic parameters. For example, 
locating experimentally the crossing point of the two eigenmodes (see
Fig. 1) would fix the value of the $\zeta$ parameter, 
provided the force component proportional to $\zeta'$ is
negligible. Measuring in addition the position of the centroid around
which the modes are split  in the strong-coupling regime ($\zeta \gg
1$) would fix both  $\zeta$ and $\zeta'$ parameters. The magnitude of
the splitting in the strong-coupling regime, 
provides information on the  volume filling factor and
the fractional densities of the condensate and the cloud.
If the Iordansky force is operative the damping of the modes in the 
strong-coupling regime would be significantly affected (asymptotically 
constant, instead of going to zero, see Fig. 1).

{\it second-order harmonic oscillations.} These oscillations 
are more complex, since they involve a larger number of modes and 
the center of mass and relative oscillations can be coupled.
The features of the 
relative transverse-shear ($m=\pm1$) and toroidal ($m=\pm2$) 
modes are as follows.  
The condition $\tilde \beta = 1$ defines a 
mode crossing point for the high-frequency modes.
The real parts of the modes  are asymptotically 
constant and the imaginary parts tend
to zero as $\zeta\to\pm \infty$,
except when $\zeta' \sim \zeta$ and
$\zeta \to \infty$, in which case  the imaginary parts tend to constants.
The bell-shaped form of the imaginary parts with maxima at 
${\rm log}_{10}~\zeta = 0 $ and  ${\rm log}_{10}~\zeta = 0.4$ (for $\zeta'=0$)
is generic to the relative   transverse-shear ($m=\pm1$) 
and toroidal ($m=\pm2$) modes. 
Increasing the rotation frequency enhances the 
damping of the modes (left/right panels of Figs. 3-5)  
since the dissipation scales linearly with the spin frequency.
The asymptotic splitting of the eigenmodes in the strong- and 
weak-coupling limits is enhanced linearly in $\Omega$. The distinctive
feature of the second-order harmonic  
pulsation modes ($m=0$) is that they couple the relative and
center of mass oscillations even when the viscosity of the matter is
negligible. There is no mode crossing for the pulsation modes
(for the common case $\zeta'=0$) and the damping has its maxima 
at $\zeta=0$ only. 

{\it Implications for experiments.}
The drag-to-lift ratios depend on the elementary processes of 
quasi-particle--vortex scattering, the strength of the interparticle 
interactions, and on thermodynamic parameters, such as the density 
and the temperature. Note that the $l=0, \pm 2 $ modes of a rotating Bose gas
have been measured experimentally at zero-temperature and are
consistent with results quoted in the Introduction.
A possible scenario of mapping out the dependence of the modes on the
mutual friction parameters could involve 
tuning the interparticle interactions in the vicinity of a Feshbach
resonance and sizably changing the quasi-particle--vortex scattering
cross section by small variations in the magnetic field. 
Varying the thermodynamic parameters, for example, the
temperature of the system is another possibility.
If the system is prepared (or the  bosonic species 
are chosen) such that the strong-coupling regime is operative,  measuring the
relative oscillations modes at various spin frequencies 
can provide information on the actual  values of the mutual 
friction and related parameters.  The mutual
friction coefficients can also be fixed  by measuring the damping of the
modes, in particular the strong-coupling limit discriminates the case 
where the Iordansky force is operative.
Experimental studies of the dependence of the
second-order harmonic modes on the lift-to-drag ratios and the
variations of the spin-frequency, apart from fixing  
the parameters reflecting the  microphysics of
quasi-particle--vortex interactions can, in addition, provide information
on the magnitude of the kinematic viscosity  from the damping
or splitting of the eigenfrequencies of the center of mass modes. 
We remind that  Eqs. (\ref{int1})-(\ref{int4})  for the
center of mass modes remain valid at finite temperatures whenever
these are decoupled from the relative oscillations 
(e.g. the $\nu \to  0$ limit).

{\it Other systems.}
The method described above could be
applied to superfluid Fermi systems at finite temperatures with 
minor modifications. For 
Fermi superfluids which rotate as a rigid body and are confined 
by a harmonic trap, the oscillation modes will differ from those
described above to the extent the equations of states of the normal 
and superfluid  differ from those of a Bose gas. Since the
linearized perturbation equations involve pressure perturbations 
only for the pulsation modes, the remainder oscillations would 
be identical to those described above. The pulsation modes can be 
easily handled by using an appropriate adiabatic index for a 
Fermi-superfluid at finite temperatures.

\appendix
\section{Virial equations for uniform rotations}

Taking the moment of Eq. (\ref{eq:euler}) with
weight 1 and integrating over $V_{\alpha}$ we obtain the first order
virial equation
\baray
&&{d\over dt}\int_{V_{\alpha}}{d^3x\rho_{\alpha} u_{_{\alpha},i}}
=2\epsilon_{ilm}\Omega_m\int_{V_{\alpha}}{d^3x\rho_{\alpha}
u_{_{\alpha}l}}\nonumber\\
&&+(\Omega^2\delta_{ij}-\omega_i^2)I_{\alpha,i}
-\Omega_i\Omega_jI_{\alpha,i}
+\int_{V_{\alpha}}d^3x F_{\alpha\beta,i},
\label{eq:v1}
\earay
where 
\be 
I_{\alpha,i}=\int_{V_{\alpha}}{d^3x\rho_{\alpha} x_i}
\ee
is the moment of inertia of fluid $\alpha$,
and we used a boundary condition 
which assumes that the projection of the stress 
$$
-p_{\alpha}\delta_{ik}+\delta_{\alpha, N}P_{ik},\quad \alpha = S, N
$$
orthogonal to a bounding surface vanishes in equilibrium (the
free-surface condition).

Taking the first moment of Eq. (\ref{eq:euler}) with weight $x_i$ and 
integrating over the volume $V_{\alpha}$ results in the
second-order virial equation
\baray
&&{d\over dt}\int_{V_{\alpha}}{d^3x\rho_{\alpha} x_j u_{\alpha,i}}
=
2{\cal  T}_{{\alpha},ij}+\delta_{ij}\Pi_{\alpha}+\delta_{\alpha, N}
{\cal P}_{ij}\nonumber\\
&&+(\Omega^2 - \omega_i^2) I_{{\alpha},ij}
+2\epsilon_{ilm}\Omega_m
\int_{V_{\alpha}}{d^3x\rho_{\alpha} x_ju_{\alpha,l}}\nonumber\\
&&
-\Omega_i\Omega_kI_{{\alpha},kj}+\int_{V_{\alpha}}d^3x x_jF_{\alpha\beta,i},
\label{eq:v2}
\earay
where
\baray
&&I_{{\alpha},ij}\equiv \int_{V_{\alpha}}{d^3x\,\rho_{\alpha} x_ix_j}, \quad
\delta_{ij}\Pi_{\alpha} \equiv \delta_{ij}
\int_{V_{\alpha}}d^3x\,p_{\alpha}, \nonumber\\
&&{\cal T}_{{\alpha},ij}\equiv {1\over 2}\int_{V_{\alpha}}{d^3x\rho_{\alpha}
u_{\alpha,i}u_{\alpha,j}}, \quad
{\cal P}_{ij} = \int_{V_N}d^3x P_{ij},\nonumber\\
\earay
are the second rank tensors of  moment of inertia, pressure, 
kinetic energy and stress; note that the last tensor is nonzero only 
in the volume of the thermal cloud and the integration over the mutual 
friction force $F_{\alpha,\beta}$ in Eq. (\ref{eq:v2}) is over the
overlap volume of the normal cloud and the condensate.

For uniformly rotating  condensate and thermal 
cloud the stationary second-order
virial equations define the equilibrium forms of the rotating system; 
we find
\baray
\Omega^2(I_{\alpha ,ij}-\delta_{i3}I_{\alpha ,3j}) 
-\omega_i^2 I_{\alpha ,ij} &=&-\delta_{ij}\Pi_{\alpha}, \quad \alpha =
S,N, \nonumber\\
\earay 
where the mixture rotates about the $x_3$ axis. 
The diagonal components of these equations provide the equilibrium 
ratios of the components of the moment of inertia tensor
\baray\label{RATIO_MI1}
\frac{I_{N,33}}{I_{N,ii}} =\frac{I_{S,33}}{I_{S,ii}} 
= \frac{\omega_i^2-\Omega^2}{\omega_3^2}, \quad i = 1,2.
\earay
The moment of inertia tensor of a heterogeneous ellipsoid is
\be\label{MOM_IN_HET}
I_{\alpha, ij} = \frac{4\pi}{3}a_{\alpha,i}^3 a_{\alpha,k}a_{\alpha,l}\delta_{ij}
\int_0^1 \rho_{\alpha}(m_{\alpha}^2)m_{\alpha}^4 dm_{\alpha},
\ee
where the density distributions $\rho_{\alpha}(m^2_{\alpha})$ are
defined in Eqs. (\ref{profile_S}) and (\ref{profile_N}).
The total mass in the fluid $\alpha$, $M_{\alpha}$ 
can be related to the moment of  inertia tensor by noting that 
\be\label{MASS_HET}
M_{\alpha} = 4\pi a_{\alpha 1}a_{\alpha 2}a_{\alpha 3}
\int_0^1\rho_{\alpha}(m_{\alpha}^2)m_{\alpha}^2dm_{\alpha}.
\ee
Using Eq. (\ref{MOM_IN_HET}) in Eq. (\ref{RATIO_MI1})
we find the ratios of the semi-axis in equilibrium 
\baray\label{RATIO_SEMIAXIS_S}
\frac{a_{N,3}}{a_{N,i}} = \frac{a_{S,3}}{a_{S,i}} =
\frac{\omega_i}{\omega_3}
\left(1-\frac{\Omega^2}{\omega^2_i} \right)^{1/2},\quad i = 1,2.
\earay
Despite  of the difference in the density profiles of the
condensate and the thermal cloud the semi-axis ratios (but not the
semiaxis!) are equal;
they depend only on the rotation rate and hence are the same  
for corotating fluids.
For axially symmetric figures these conditions place  
an upper limit on the rotation frequency $\Omega <
\omega_{\perp}$. In the nonaxisymmetric case $\Omega < {\rm min}
(\omega_1,\,\omega_2)$ for fixed $\omega_1$ and $\omega_2$ 
or, alternatively,  $\omega_2> \Omega$ for fixed $\omega_{1}$ 
at $\Omega<\omega_1$. Note that these boundaries correspond
to stable solutions under stationary conditions and within the ellipsoidal
approximation. Perturbations away from the stability region can
result in either dynamically unstable figures that preserve their
ellipsoidal structure or (stable) configurations that are not
ellipsoids (e.g., the Poincar\'e's pear-shaped figures) \cite{CHANDRA}.

\section{Variations of the stress energy and pressure tensors}

The variations of the stress tensor, Eq. (\ref{eq:stress_tensor}),
cannot be expressed in terms of virials in general. 
This can be done, however, in the perturbative regime, when the effects 
of viscosity are small.
The perturbation expansion about  the inviscid limit requires 
$\sigma_1\ll\sigma_0$, where $\sigma_1$ and $\sigma_0$ are the modes 
in the viscous and inviscid limits. Since the characteristic equations 
which include fluid viscosity do not permit analytical solutions in
general, the validity of the perturbation expansion can be checked 
numerically ({\it a posteriori}) by comparing the values of $\sigma_0$ and 
$\sigma_1$ for a fixed mode. Another expansion 
parameter, that enters only the equations for the relative virials is 
$\nu/\beta\Omega\ll 1$; the latter condition insures that the
modifications of the damping of the modes due to the viscosity are 
small compared to the inviscid but dissipative (due to the mutual
friction) case.
The variation of the stress energy tensor
in the background equilibrium without internal motions ($\uvec = 0$) 
is 
\be\label{PIVAR} 
\delta {\cal P}_{ij} = \nu\int_{V_N}\rho_N \frac{\partial}{\partial t}
\left(\frac{\partial\xi_{N,j}}{\partial x_i}
+\frac{\partial \xi_{N,i}}{\partial x_j}
-\frac{2}{3}\frac{\partial\xi_{N,l}}{\partial x_l}\delta_{ij} \right)d^3x,
\ee 
Eq. (\ref{PIVAR}) assumes that the kinematic viscosity 
can be replaced by its average over the profile of the normal 
cloud, in line with the similar assumption about the mutual 
friction coefficients.

The perturbative approach expresses the stress energy tensor 
in terms of Lagrangian displacements corresponding to the 
inviscid limit (see, e.g., \cite{CHANDRA}). 
For perturbations which are linear in the coordinates, these can
be written as
\be\label{XI_INVISC} 
\xi_{N,i}= L_{N,i;j}x_j
\ee
where $L_{N,i;j}$ are nine unknowns which are determined from the
solution of the virial equations of second-order in the inviscid limit
[no summation over the repeated indices in Eq. (\ref{XI_INVISC})]. 
Using Eqs. (\ref{eq:XI_T}), (\ref{XI_INVISC}) in (\ref{PIVAR}) we obtain
\be\label{PIVAR2} 
\delta {\cal P}_{ij} = -i\sigma\nu\int_{V_N}\rho_N 
\left(L_{N,i;j}+L_{N,j;i}-\frac{2}{3}\delta_{ij}L_{N,l;l} \right)d^3x.
\ee
On the other hand, noting that the moment of inertia is diagonal
($I_{ij}=0$, for $i\neq j$) due to the tri-planar geometry of the
ellipsoid
\be\label{eq:V_L}
V_{N,i;j} \equiv \int_{V_N}\rho_N~\xi_{N,i}x_j d^3x = L_{N,i;j}I_{N,jj},
\ee
(no summation over the repeated indices). Next using the expressions for 
the moment of inertia tensor and the mass of heterogenous ellipsoids,
Eqs. (\ref{MOM_IN_HET}) and (\ref{MASS_HET}), we write the ratio
$I_{N,ii}/M_N = \bar a_{i}^2/5$, where
\be\label{EFF_SEMIAXIS} 
\bar a_{i}^2 \equiv 5a_{N,i}^2 
{\int_0^1\rho_N(m_N^2)m_N^4dm_N}
\left[3{\int^1_0\rho_N(m_N^2)m_N^2dm_N}\right]^{-1}.
\ee
Then, Eq. (\ref{PIVAR2}) becomes
\be 
\delta {\cal P}_{ij} = -5i\nu\sigma\left(
\frac{V_{N,i;j}}{\bar a_j^2}
+\frac{V_{N,j;i}}{\bar a_i^2}
-\delta_{ij}\frac{V_{N,ll}}{3\bar a_l^2}
\right).
\ee  
The final form of the equations governing the small-amplitude
oscillations in the viscous fluid limit will contain only the ratios of 
the semi-axis of the ellipsoids; 
therefore instead of evaluating the expression 
(\ref{EFF_SEMIAXIS}),  the relations (\ref{RATIO_SEMIAXIS_S}) will be
 used. 

As pointed out in the main text, the center of mass and relative quasi-radial 
oscillations modes are coupled due to the difference in the
perturbations of the stress tensor of the condensate and the 
thermal cloud. We now express these perturbations in terms of
the virials $V_{ij}$ and $U_{ij}$. The Eulerian perturbation for 
the condensate has been derived earlier [Paper I, Eq. (42)]
\be\label{PI_S}  
\delta\Pi_S =
\frac{1-\gamma}{2}\left[(\omega_{\perp}^2-\Omega^2)(V_{S,11}+V_{S,22})
+\omega_3^2 V_{S,33} \right],
\ee
where $\gamma = 2$ for the condensate.
Adiabatic perturbations of the thermal cloud leave the quantity 
$p_N\rho_N^{-5/3}$ unchanged. For the perturbation of the stress
tensor of the thermal cloud we find
\be\label{PI_N1} 
\delta\Pi_N = \int_{V_N} (\Delta p_N + p_N \bnabla\cdot \xivec_N )d^3x
 = \frac{2}{3} \int_{V_N} \xivec_N\cdot  \bnabla p_N~d^3x,
\ee
where we used the relation between the Eulerian ($\delta$) and
Lagrangian ($\Delta$) variations: $\delta = \Delta +  \bnabla\cdot
\xivec$. The gradient of the pressure $p_N$ is computed from the 
equilibrium limit of the Navier-Stokes equation (\ref{eq:euler}). We find
\be\label{PI_N2}  
\delta\Pi_N =
\frac{1-\gamma'}{2}\left[(\omega_{\perp}^2-\Omega^2)(V_{N,11}+V_{N,22})
+\omega_3^2 V_{N,33} \right],
\ee
where the effective adiabatic index is $\gamma'= 5/3$. Note that the
adiabatic index governing the perturbations need not be identical to
the one that governs the equilibrium background, as is the 
case for the thermal cloud. The explicit expression for the Eulerian 
variations are obtained upon using Eqs. (\ref{PI_S}) and (\ref{PI_N2})
\baray \label{Pi_plus}
\delta\Pi^{(+)} &=& \frac{1-\gamma'+f(1-\gamma)}{2(1+f)}\nonumber\\
&\times&
\left[(\omega^2_{\perp}-\Omega^2)(V_{11}+V_{22})+\omega_3^2V_{33}\right]
\nonumber\\
&+&\frac{\gamma'-\gamma}{2(1+f)}
\left[(\omega^2_{\perp}-\Omega^2)(U_{11}+U_{22})+\omega_3^2U_{33}\right],
\nonumber\\
\\
\label{Pi_minus}
\delta\Pi^{(-)} &=& \frac{f(\gamma'-\gamma) }{2(1+f)}
\left[(\omega^2_{\perp}-\Omega^2)(V_{11}+V_{22})+\omega_3^2V_{33}\right]
\nonumber\\
&+&\frac{1-\gamma+f(1-\gamma')}{2(1+f)}\nonumber\\
&\times&
\left[(\omega^2_{\perp}-\Omega^2)(U_{11}+U_{22})+\omega_3^2U_{33}\right].
\earay


\begin{thebibliography}{99}
\bibitem{FOOTNOTE} 
To rewrite the expressions of Paper I 
in the present notations substitute 
$\omega_0^2A_3 \equiv \omega_{\perp}^2 A_3 \to\omega_z^2$ and specify
                    Eq. (45) for the case of the adiabatic 
                    index value $\gamma=2$. Note that the modes in 
                    Paper I are given in a dimensionless form, the 
                    frequency unity being $\omega_{\perp}$; and a normalization
                    is used, $A_1=1$, such that $\omega_{\perp}$ is identical 
                    to the fundamental frequency $\omega_0.$
\end{thebibliography}
\end{document}